\documentclass[12pt]{article}
\usepackage[dvips]{graphicx}
\begin{document}
\title{WHAT CAN WE LEARN FROM THE STUDY 
OF SINGLE DIFFRACTIVE DISSOCIATION AT HIGH ENERGIES?\footnote{The
talk presented at the VIIIth Blois Workshop on Elastic and
Diffractive Scattering. Protvino, Russia, June 28--July 2, 1999.}}
\author{ {A.A.~Arkhipov}\\
\it{Institute for High Energy Physics}\\
\it{Protvino, 142284 Moscow Region, Russia}}
\date{}
\maketitle
\def\be{\begin{equation}}
\def\ee{\end{equation}}
\def\ber{\begin{eqnarray}}
\def\eer{\end{eqnarray}}
\begin{abstract}
The fundamental relations in the dynamics of single diffraction
dissociation and elastic scattering at high energies are discussed.
\end{abstract}


 
\section{Introduction}
I shall start my talk from the end with the short answer to the
question in the title. The study of the inclusive
reactions in the region of diffraction dissociation at high energies
provides a unique possibility to learn on a new type of interactions
between elementary particles or a new type of fundamental forces,
which the three-body forces are. What was the beginning on?

In 1994 the CDF group at Fermilab published new results on the
measurements of $p\bar p$ single diffraction dissociation 
at $\sqrt{s} = 546$ and $1800\ GeV$. They observed that a popular
supercritical Pomeron model did not describe new measured values.
The statement, made in  \cite{1}, is as follows: The value of
$\sigma_{sd}^{p\bar p} = (7.89\pm 0.33)\ mb$, measured at $\sqrt{s} =
546\ GeV$,
is extrapolated by the supercritical Pomeron model to
$\sigma_{sd}^{p\bar p} = (13.9\pm 0.9)\ mb$ at $\sqrt{s} = 1800\
GeV$, while the measured value at this energy is equal to
$\sigma_{sd}^{p\bar p} =
(9.45\pm 0.44)\ mb$. The ratio of the measured $\sigma_{sd}^{p\bar
p}$ to that obtained by extrapolation is
\be
\frac{\sigma_{sd}^{p\bar p}(experimental)}{\sigma_{sd}^{p\bar
p}(extrapolation)}
(\sqrt{s} = 1800 \ GeV) = 0.68\pm 0.05 . \label{1}
\ee
Moreover, at $\sqrt{s} = 20 \ GeV$ the experimental
$\sigma_{sd}^{p\bar p} = (4.9\pm 0.55)\ mb$ is 4.5 times larger than
the value $\sigma_{sd}^{p\bar p} = (1.1\pm 0.17)\ mb$, obtained by
the extrapolation of the measured value of $\sigma_{sd}^{p\bar p}$ at
$\sqrt{s} = 546\ GeV$ down to $\sqrt{s} = 20\ GeV$ with the help of
the supercritical Pomeron model. So, the latest experimental
measurement of $p\bar p$ single diffraction dissociation at c.m.s.
energies $\sqrt{s} = 546$ and $1800\ GeV$, carried out by the CDF
group
at the Fermilab Tevatron collider,  has shown that the popular model
of supercritical Pomeron does not describe the existing experimental
data.

We called the emerged situation as a supercrisis for the
supercritical Pomeron model (SCPM).\footnote{Recent experimental
results from HERA \cite{2} lead us to the same conclusion. The soft
Pomeron phenomenology as currently developed cannot incorporate the
HERA data on structure function $F_2$ at small $x$ and total
$\gamma^{*}p$ cross section from $F_2$ measurements as a function of
$W^2$ for different $Q^2$.} The supercrisis is illustrated on Fig. 1
extracted from paper \cite{3}.

The attempts undertaken in Refs. \cite{3,4} to save the SCPM are
also shown on this figure. Unfortunately GLM paper \cite{4} contains
a crude mathematical mistake. The mistake was observed by B.V.
Struminsky and E.S. Martynov from Kiev \cite{5}. Besides, in our
opinion, an eikonalization procedure cannot be considered as a saving
ring for SCPM because this procedure is outside the original Regge
ideology. The idea of renormalized Pomeron flux proposed by Goulianos
is a good physical idea for an experimentalist, but this idea cannot
be a satisfactory one for a theorist because the idea is not
grounded by the underlying Regge theory.

Obviously, the foundations of the Pomeron model require a further
theoretical study and the construction of newer, more general
phenomenological framework, which would enable one to remove the
discrepancy between the model predictions and the experiment.

Although nowadays we have in the framework of local quantum field
theory a gauge model of strong interactions formulated in terms of
the known QCD Lagrangian, its relations to the so called ``soft"
(interactions at large distances) hadronic physics are far from
desired. The understanding of this physics is high interest because
it has an intrinsically fundamental nature.

In 1970 the experiments at the Serpukhov accelerator revealed that
the $K^+p$ total cross section increased with energy. The increase of
the $pp$ total cross section was discovered at the CERN ISR and
then the effect of rising total cross sections was confirmed at the
Fermilab accelerator.

In spite of more than 25 years after the formulation of QCD we still
cannot obtain from the QCD Lagrangian the answer to the question why
all the hadronic total cross-sections grow with energy. We cannot
predict total cross-sections in an absolute way starting from the
fundamental QCD Lagrangian as well mainly because it is not a
perturbative problem.

It is well known, e.g., that nonperturbative contributions to the
gluon propagator influence the behaviour of ``soft" hadronic
processes and the knowledge of the infrared behaviour of QCD is
certainly needed to describe the ``soft" hadronic physics in the
framework of QCD. Unfortunately, today we don't know the whole
picture of the infrared behaviour of QCD, we have some fragments of
this picture though (see e.g. Ref. \cite{6}).

At the same time it is more or less clear now that the rise of the
total cross-sections is just the shadow (not antishadow!) of particle
production. 

Through the optical theorem the total cross-section is related to the
imaginary part of the elastic scattering amplitude in the forward
direction. That is why the theoretical understanding of elastic
scattering has the fundamental importance.

From the unitarity relation it follows that the imaginary part of the
elastic scattering amplitude contains the contribution of all
possible inelastic channels in two-particle interaction. It is clear
therefore that we cannot understand the elastic scattering without
understanding the inelastic interaction.

Among all the possible inelastic interactions there is a special
class of processes which are called a single diffraction
dissociation. The single diffraction dissociation is the scattering
process where one of two particles in the initial state breaks up
during the interaction producing a system of particles in a limited
region of (pseudo)rapidity.\footnote{Pseudorapidity is defined as
$\eta = -\ln \tan (\theta /2)$ where $\theta$ is the  polar angle of
the produced particle with respect to the beam direction.
Pseudorapidity is frequently used as an approximation to rapidity.}

Good and Walker have shown \cite{7} that the single diffraction
dissociation is predicted by the basic principles of quantum
mechanics. However both the elastic scattering and single diffraction
dissociation cannot correctly be calculated in QCD due to the
non-perturbative nature of the interactions.

The popular Regge phenomenology represents elastic
and diffractive scattering by the exchange of the Pomeron, a color
singlet Reggeon with quantun numbers of the vacuum. It should be
noted that the definition of the Pomeron as Reggeon with the highest
Regge trajectory $\alpha_P(t)$, carrying the quantum numbers of
the vacuum, is not the only one.\footnote{For supercritical Pomeron
$\alpha_P(0) - 1 = \Delta \ll 1,\,  \Delta > 0$ is responsible for
the growth of hadronic cross-sections with energy.} There are many
other definitions  of the Pomeron: Pomeron is a gluon ``ladder" 
\cite{8}; Pomeron is a bound state of two reggezied gluons --
BFKL-Pomeron \cite{9}; soft and hard Pomerons \cite{10,11};
etc.\footnote{At the Workshop I heard new definition of Pomeron from
N.N. Nikolaev: Pomeron is (neither more nor less!) a label of
diffraction.} This leapfrog is because of the exact nature of the
Pomeron and its detailed substructure remains such as that no one
knows what it is. The difficulty of establishing the true
nature of the Pomeron in QCD is almost obviously related to the
calculations of non-perturbative gluon exchange.

Nevertheless in the near past simple formulae of the Regge
phenomenology provided good parameterization of experimental data on
``soft" hadronic physics and pragmatic application of Pomeron
phenomenology had been remarkably successful (see e.g. the latest
issue of the Review of Particles Properties).

That was the case before the appearance of the above-mentioned CDF
data on single diffractive dissociation and recent results from HERA.
Of course, it is good that we have a simple and compact form for
representing a great variety of data for different hadronic
processes, but it is certainly bad that power behaved total
cross-sections violate unitarity. Often and often encountered claim,
that
the model with power behaved total cross-sections is valid in the
non-asymptotic domain which has been explored up today, is not
correct because the supercritical Pomeron model is an asymptotic one
by definition.

We suggested another approach to the dynamical description of
one-particle inclusive reactions \cite{12}. The main point of our
approach is that new fundamental three-body forces are responsible
for the dynamics of particle production processes of inclusive type.
Our consideration revealed several fundamental properties of
one-particle inclusive cross-sections in the region of diffraction
dissociation. In particular, it was shown that the slope of the
diffraction cone in $p\bar p$ single diffraction dissociation  is
related to the effective radius of three-nucleon forces in the same
way as the slope of the diffraction cone in elastic $p\bar p$
scattering is related to the effective radius of two-nucleon
forces. It was also demonstrated that the effective radii of two- and
three-nucleon forces, which are the characteristics of elastic and
inelastic interactions of two nucleons, define the structure of the
$p\bar p$ total cross-sections in a simple and physically clear
form. I'll touch upon these properties later on.

First of all let me tell you a few words what I mean by three-body
forces about.

\section{Three-body forces in relativistic quantum theory}

Using the LSZ or the Bogoljubov reduction formulae in quantum field
theory \cite{13} we can easily obtain the following cluster structure
for $3\rightarrow 3$ scattering amplitude (see diagram below)
\be
{\cal F}_{123} = {\cal F}_{12} + {\cal F}_{23} + {\cal F}_{13} + 
{\cal F}_{123}^C \label{2}
\ee
where ${\cal F}_{ij} , (i,j = 1,2,3)$ are $2 \rightarrow 2$
scattering  amplitudes, ${\cal F}_{123}^C$ is called the connected
part of the $3 \rightarrow 3$ scattering amplitude.

\vskip0.4cm
\begin{center}
\begin{picture}(370,30)
\thicklines
\put(0,0){\line(1,0){60}}
\put(0,30){\line(1,0){60}}
\put(30,15){\circle{30}}
\put(0,15){\line(1,0){15}}
\put(45,15){\line(1,0){15}}

\put(65,12){\mbox{=}}

\put(75,0){\line(1,0){60}}
\put(75,15){\line(1,0){60}}
\put(105,22,5){\circle{15}}
\put(75,30){\line(1,0){60}}

\put(140,12){\mbox{+}}

\put(150,0){\line(1,0){60}}
\put(150,15){\line(1,0){60}}
\put(150,30){\line(1,0){60}}
\put(180,7.5){\circle{15}}

\put(215,12){\mbox{+}}

\put(230,0){\line(1,0){60}}
\put(230,30){\line(1,0){60}}
\put(260,15){\circle{30}}
\put(260,15){\oval(40,20)[b]}
\put(230,15){\line(1,0){10}}
\put(280,15){\line(1,0){10}}

\put(300,12){\mbox{+}}

\put(315,0){\line(1,0){60}}
\put(315,30){\line(1,0){60}}
\put(345,15){\circle{30}}
\put(315,15){\line(1,0){15}}
\put(360,15){\line(1,0){15}}

\put(335,7){\mbox{\huge C}}

\end{picture}
\end{center}

In the framework of single-time formalism in quantum field theory
\cite{14} we construct the $3 \rightarrow 3$ off energy shell
scattering amplitude $T_{123}(E)$ with the same (cluster) structure 
as (\ref{2})
\be
T_{123}(E) = T_{12}(E) + T_{23}(E) + T_{13}(E) + T_{123}^C(E).
\label{3}
\ee
Following the tradition we'll call the kernel describing the
interaction of three particles as the three particle interaction
quasipotential. The three particle interaction quasipotential
$V_{123}(E)$ is related to the off-shell $3 \rightarrow 3$ scattering
amplitude $T_{123}(E)$by the Lippmann-Schwinger type equation
\be
T_{123}(E) = V_{123}(E) + V_{123}(E)G_0(E)T_{123}(E). \label{4}
\ee
There exists the same transformation between two particle
interaction quasi\-potentials $V_{ij}$ and off energy shell $2
\rightarrow 2$ scattering amplitudes $T_{ij}$
\be
T_{ij}(E) = V_{ij}(E) + V_{ij}(E)G_0(E)T_{ij}(E). \label{5}
\ee
It can be shown that in the quantum field theory the three 
particle interaction quasipotential has the following structure
\cite{15}
\be
V_{123}(E) = V_{12}(E) + V_{23}(E) + V_{13}(E) + V_0(E). \label{6}
\ee
The quantity $V_0(E)$ is called the three-body forces quasipotential.
The $V_0(E)$ represents the defect of three particle interaction
quasipotential over the sum of two particle interaction
quasipotentials and describes the true three-body interactions.
The three-body forces quasipotential is an inherent connected part of
total three particle interaction quasipotential which cannot be
represented by the sum of pair interaction quasipotentials.

The three-body forces scattering amplitude is related to the
three-body forces quasipotential by the equation
\be
T_0(E) = V_0(E) + V_0(E)G_0(E)T_0(E). \label{7}
\ee

It should be stressed that the three-body forces appear as a 
result of consistent consideration of three-body problem in the
framework of local quantum field theory.

\section{Global analyticity of the three-body forces} 

Let us introduce the following useful notations
\be
<p'_1 p'_2 p'_3\vert S - 1\vert p_1 p_2 p_3> =
2\pi i\delta ^4(\sum_{i=1}^{3}p'_i-\sum_{j=1}^{3}p_j)
{\cal F}_{123}(s;{\hat e}',\hat e), \label{8}
\ee
\[
s = (\sum_{i=1}^{3}p'_i)^2 = (\sum_{j=1}^{3}p_j)^2.
\]
The ${\hat e}', \hat e \in S_5$ are two unit vectors on
five-dimensional sphere describing the configuration of three-body
system in the initial and final states (before and after scattering).

We will denote the quantity $T_0$ restricted on the energy shell as
\[
T_0\mid_{on\, energy\, shell}\, = {\cal F}_0.
\]
The unitarity condition for the quantity ${\cal F}_0$ with 
account for the introduced notations can be written in form
\cite{16,17}
\ber
\lefteqn{Im{\cal F}_0(s;{\hat e}',\hat e)=}\nonumber\\
&=& \pi A_3(s)\int d\Omega _{5}({\hat e}''){\cal F}_0(s;{\hat
e}',{\hat e}'')
\stackrel{*}{\cal F}_0(s;\hat e,{\hat e}'') + H_0(s;{\hat e}',\hat
e),
\label{9}
\eer

\[
Im{\cal F}_0(s;{\hat e}',\hat e)=\frac{1}{2i}\left[{\cal F}_0(s;{\hat
e}',\hat e)
-\stackrel{*}{\cal F}_0(s;\hat e,{\hat e}')\right],
\]
where
\[
A_3(s) = {\Gamma}_3(s)/S_5 ,
\]
${\Gamma}_3(s)$ is the three-body phase-space volume, $S_5$ is the
volume of unit five-dimensional sphere. $H_0$ defines the
contribution of all the inelastic channels emerging due to three-body
forces.

Let us introduce a special notation for the scalar product of two
unit vectors 
${\hat e}'$ and $\hat e$
\be
\cos \omega = {\hat e}^{'}\cdot \hat e. \label{10}
\ee
We will use the other notation for the three-body forces scattering
amplitude as well 
\[
{\cal F}_0(s;{\hat e}',\hat e) = {\cal F}_0(s;\eta,\cos\omega),
\]
where all other variables are denoted through $\eta$.

Now we are able to go to the formulation of our basic assumption
on the analytical properties for the three-body forces scattering
amplitude 
\cite{16,17}.

We will assume that for physical values of the variable $s$ and fixed 
values of  $\eta$ the amplitude ${\cal F}_0(s;\eta,\cos\omega)$ is an 
analytical function of the variable $\cos\omega$ in the ellipse
$E_0(s)$ 
with the semi-major axis
\be
z_0(s) = 1 + \frac{M_0^2}{2s} \label{11}
\ee
and for any $\cos\omega \in E_0(s)$ and physical values of $\eta$ it
is polynomially bounded in the variable $s$. $M_0$ is some constant
having mass dimensionality.

Such analyticity of the three-body forces amplitude was called a
global one. The global analyticity may be considered as a direct
geometric generalization of the known analytical properties of
two-body scattering amplitude strictly proved in the local quantum
field theory \cite{18,19,20,21,22}.

At the same time the global analyticity results in the generalized
asymptotic bounds. 
\begin{center}
$\framebox{\bf GLOBAL ANALYTICITY}\enspace \& \enspace \framebox{\bf
UNITARITY}$\\
\vskip 2ex
$\Downarrow$\\
\vskip 2ex
$\framebox{\bf GENERALIZED ASYMPTOTIC BOUNDS}$
\end{center}

For example the generalized asymptotic bound for
$O(6)$-invariant three-body forces scattering amplitude looks like
\cite{16,17}
\begin{equation}
Im\,{\cal F}_0(s;...) \leq \mbox{Const}\, s^{3/2} 
\bigl(\frac{\ln s/s'_0}{M_0}\bigr)^5 = \mbox{Const}\,
s^{3/2}R_0^5(s),
\label{12}
\end{equation}
where $R_0(s)$ is the effective radius of the three-body forces
introduced according to  \cite{22} where the effective radius
of two-body forces has been defined,
\be
R_0(s) = \frac{\Lambda_0}{\Pi(s)} = \frac{r_0}{M_0}\ln
\frac{s}{s'_0},
\quad \Pi(s) = \frac{\sqrt{s}}{2},\quad s\rightarrow \infty,
\label{13}
\ee
$r_0$ is defined by the power of the amplitude ${\cal F}_0$ growth 
at high energies \cite{17}, $M_0$ defines the semi-major axis of the
global analyticity ellipse (\ref{11}), ${\Lambda}_0$ is the effective
global orbital momentum, $\Pi(s)$ is the global momentum of
three-body system, $s'_0$ is a scale defining unitarity saturation of
three-body forces.
 
It is well known that the Froissart asymptotic bound \cite{23} can
be experimentally verified, because with the help of the optical
theorem we can  connect the imaginary part of $2 \rightarrow 2$
scattering amplitude with the experimentally measurable quantity
which is the total cross-section. So, if we want to have a
possibility for the experimental verification of the generalized
asymptotic bounds $(n\geq3)$, we have to establish a connection
between the many-body forces scattering  amplitudes and the
experimentally measurable quantities. For this aim we have considered
the problem of high energy particle scattering from deuteron and  on
this way we found the connection of the three-body forces scattering
amplitude with the experimentally measurable quantity which is the
total cross-section for scattering from deuteron \cite{24}. Moreover
the relation of the three-body forces scattering amplitude to
one-particle inclusive cross-sections has been established \cite{25}. 

I shall briefly sketch now the basic results of our analysis of
high-energy particle scattering from deuteron. 

\section{Scattering from deuteron}
The problem of scattering from two-body bound states was treated in
\cite{24,25} with the help of dynamic equations obtained on the
basis of single-time formalism in QFT \cite{15}. As has been shown in
\cite{24,25}, the total cross-section in the scattering from deuteron
can  be expressed by the formula
\be
\sigma_{hd}^{tot}(s) = \sigma_{hp}^{tot}(\hat s)
+\sigma_{hn}^{tot}(\hat s) -
\delta\sigma(s), \label{14}
\ee
where $\sigma_{hd}, \sigma_{hp}, \sigma_{hn}$  are the total
cross-sections
in scattering from deuteron, proton and neutron, 
\begin{equation}
\delta\sigma(s) = \delta\sigma_G(s) +\delta\sigma_0(s),\label{15}
\end{equation}
\begin{equation}
\delta\sigma_G(s) =
\frac{\sigma_{hp}^{tot}(\hat s) \sigma_{hn}^{tot}(\hat s)}{4\pi(
R^2_d+B_{hp}(\hat s)+B_{hn}(\hat s))} 
\equiv \frac{\sigma_{hp}^{tot}(\hat s) \sigma_{hn}^{tot}(\hat
s)}{4\pi
R^2_{eff}(s)},\ \ \hat s = \frac{s}{2},
\label{16}
\end{equation}
$B_{hN}(s)$ is the slope of the forward diffraction peak in the
elastic scattering from nucleon, $1/R_d^2$ is defined by the deuteron
relativistic formfactor
\be
\frac{1}{R_d^2} \equiv \frac{q}{\pi }\int\frac{d\vec \Delta \Phi
(\vec \Delta )}{2\omega_h (\vec 
q+\vec \Delta )}\delta \left[\omega_h (\vec q+\vec \Delta )-\omega_h
(\vec q\,)\right],\ \ \frac{s}{2M_d}\cong q\cong \frac{\hat
s}{2M_N}, \label{17}
\ee
$\delta\sigma_G$ is the Glauber correction or shadow effect. The
Glauber shadow correction originates from elastic rescatterings of an
incident particle on the nucleons inside the deuteron.

The quantity $\delta\sigma_0$ represents the contribution of the
three-body forces to the total cross-section in the scattering from
deuteron. The physical reason for the appearance of this quantity is
directly connected with the inelastic  interactions of an incident
particle with the nucleons of deuteron. Paper \cite{25} 
provides for this quantity the following expression:
\begin{equation}
\delta\sigma_0(s) = -\frac{(2\pi)^3}{q}
\int \frac{d\vec\Delta\Phi(\vec\Delta)}
{2E_p(\vec\Delta /2)2E_n(\vec\Delta /2)}
Im\,R\bigl(s;-\frac{\vec\Delta}{2},\frac{\vec\Delta}{2},\vec q; 
\frac{\vec\Delta}{2},-\frac{\vec\Delta}{2},\vec q \bigr),\label{18}
\end{equation}
where $q$ is the incident particle momentum in the lab system (rest
frame of deuteron), $\Phi(\vec\Delta)$ is the deuteron relativistic
formfactor, normalized to unity at zero, 
\[
E_N(\vec\Delta)=\sqrt{\vec\Delta^2 + M^2_N}\quad N = p, n,
\]
$M_N$ is the nucleon mass. The function $R$ is expressed via the
amplitude of the three-body forces $T_0$ and the amplitudes of
elastic scattering from the nucleons $T_{hN}$ by the relation
\begin{equation}
R = T_0 + \sum_{N=p,n}(T_0G_0T_{hN} + T_{hN}G_0T_0).\label{19}
\end{equation}
In \cite{24} the contribution of three-body forces to the
scattering amplitude from deuteron was related to the processes of
multiparticle production in the inelastic interactions of the
incident particle with the nucleons of deuteron. This was done with
the help of the unitarity  equation. The character of the energy
dependence of $\delta\sigma_0$ was shown to be governed by the energy
behaviour of the corresponding inclusive cross-sections.

Here, for simplicity, let us consider the model where the imaginary
part of the three-body forces scattering amplitude has the form
\begin{equation}
Im\,{\cal F}_0(s; \vec p_1, \vec p_2, \vec p_3; \vec q_1, \vec q_2,
\vec q_3)
= f_0(s)
\exp \Biggl\{-\frac{R^2_0(s)}{4} \sum^{3}_{i=1} (\vec p_i-\vec
q_i)^2\Biggr\},\label{20}
\end{equation}
where $f_0(s)$, $R_0(s)$ are free parameters which, in general, may
depend on the total energy of three-body interaction. Note that the
quantity
$f_0(s)$ has the dimensionality $[R^2]$. 

In case of unitarity saturation of the three-body forces, we have 
from the generalized asymptotic theorems 
\be
f_0(s) \sim \mbox{Const}\, s^{3/2}{\Bigl(\frac{\ln
s/s'_0}{M_0}\Bigr)}^5 = \mbox{Const}\, s^{3/2}R_0^5(s),\label{21}
\ee
\be
R_0(s) = \frac{r_0}{M_0} \ln s/s'_0 \quad s\rightarrow
\infty.\label{22}
\ee

In the model all the integrals can be calculated in the
analytical form. As a result, we obtain for the quantity
$\delta\sigma_0$ \cite{25}
\[
\delta\sigma_0(s) = \frac{(2\pi)^{6}f_0(s)}{sM_N }
\Biggl\{\frac{\sigma_{hN}(s/2)}{2\pi[B_{hN}(s/2)+R^2_0(s)-R^4_0(s)/4(
R^2_
0(s)+R^2_d)]}-1\Biggr\}
\]
\be
\times\frac{1}{[2\pi(R^2_d+R^2_0(s))]^{3/2}}.\label{23}
\ee 
If the condition
\be
R_0^2(s) \simeq B_{hN}(s/2) \ll R_d^2 \label{24}
\ee
is realized, then we obtain from expression (\ref{23})
\be
\delta\sigma_0(s) =
(2\pi)^{9/2}\frac{f_0(s)\chi(s)}{sM_NR_d^3},\label{25}
\ee
where
\be
\chi(s) = \frac{\sigma_{hN}^{tot}(s/2)}{2\pi[B_{hN}(s/2)+R_0^2(s)]} -
1,\label{26}
\ee
and we suppose that asymptotically
\[
B_{hp}=B_{hn}\equiv B_{hN},\quad \sigma_{hp}^{tot}=\sigma_{hn}^{tot}
\equiv \sigma_{hN}^{tot}.
\]

It follows from the Froissart theorem and generalized asymptotic
bounds (\ref{12}) that the following asymptotic behaviour is admitted
for the $\chi(s)$:
\be 
\chi(s) \sim  \frac{1}{\sqrt{s}{\ln}^3 s},\quad s\rightarrow 
\infty. \label{27}
\ee
\section{Three-body forces in single diffraction dissociation}

From the analysis of the problem of high-energy particle
scattering from deuteron we have derived the formula connecting
one-particle inclusive cross-section with the imaginary part of the
three-body forces scattering amplitude. This formula looks like
\be 
\fbox{$\displaystyle 2E_N(\vec{\Delta})\frac{{d\sigma}_{hN\rightarrow
NX}}{d\vec{\Delta}}(s,\vec{\Delta}) = 
- \frac{(2\pi)^3}{I(s)}
Im{\cal F}_0^{scr}(\bar s;-\vec{\Delta}, \vec{\Delta}, \vec q; 
\vec{\Delta}, -\vec{\Delta}, \vec q\,)$}\,, \label{28}
\ee 
\vspace{3mm}
\[
Im{\cal F}_0^{scr}(\bar s;-\vec{\Delta}, \vec{\Delta}, \vec q; 
\vec{\Delta}, -\vec{\Delta}, \vec q\,) = Im{\cal F}_0(\bar
s;-\vec{\Delta}, \vec{\Delta}, \vec q; 
\vec{\Delta}, -\vec{\Delta}, \vec q\,)-
\]
\[
-
4\pi\int d\vec{\Delta}'\frac{\delta\left[E_N(\vec{\Delta} -
\vec{\Delta}') + \omega_h(\vec q+\vec{\Delta}') - E_N(\vec{\Delta}) -
\omega_h(\vec q)\right]}{2\omega_{h}(\vec q + \vec{\Delta}')2E_N
(\vec{\Delta} - \vec{\Delta}')}\times
\]
\be
Im{\cal F}_{hN}(\hat s; \vec{\Delta}, \vec q; 
\vec{\Delta}-\vec{\Delta}', \vec q + \vec{\Delta}'\,)Im{\cal
F}_0(\bar s;-\vec{\Delta}, \vec{\Delta}-\vec{\Delta}', \vec q +
\vec{\Delta}'; 
\vec{\Delta}, -\vec{\Delta}, \vec q\,), \label{29}
\ee 
\vspace{3mm}
\[
E_N(\vec{\Delta})=\sqrt{{\vec{\Delta}}^2+M_N^2},\quad
\omega_h(\vec q)=\sqrt{{\vec q}\,^2+m_h^2},
\]
\[
I(s) = 2{\lambda}^{1/2}(s,m_h^2,M_N^2),\quad \hat s = \frac{\bar s +
m_h^2 - 2M_N^2}{2},
\]
\[
\bar s = 2(s + M_N^2) - M_X^2,\quad t = - 4{\vec\Delta}^2.
\]
I'd like to draw the attention to the minus sign in the R.H.S. of Eq.
(\ref{28}). The simple model for the three-body forces considered
above (see Eq. (\ref{20})) gives the following result for the
one-particle inclusive cross-section in the region of diffraction
dissociation
\[
\frac{s}{\pi}\frac{d\sigma_{hN\rightarrow NX}}{dtdM_X^2} =
\frac{(2\pi)^3}{I(s)}\chi(\bar s)Im{\cal F}_0(\bar
s;-\vec{\Delta}, \vec{\Delta}, \vec q; 
\vec{\Delta}, -\vec{\Delta}, \vec q\,)
\]
\be
= \frac{(2\pi)^3}{I(s)}\chi(\bar s)f_0(\bar
s)\exp\Biggl[\frac{R_0^2(\bar s)}{2}t\Biggr] \label{30}
\ee 
where
\[
\chi(\bar s) = \frac{\sigma^{tot}_{hN}({\bar s}/2)}{2\pi[B_{hN}({\bar
s}/2) + R_0^2(\bar s)]} -1.
\]
The configuration of particles momenta and kinematical variables are
shown in Fig. 2. The variable $\bar s$ in the R.H.S. of Eq.
(\ref{30}) is related to  the kinematical variables of one-particle
inclusive reaction by the equation
\be
\bar s = 2(s + M_N^2) - M_X^2,\label{31}
\ee
\[
t = - 4{\Delta}^2.
\]

There is a temptation to call the quantity $I(s)\chi^{-1}(\bar s)$ 
a renormalized flux. However, it should be pointed out that in our
case we have a flux of real particles and function $\chi (s)$ has 
quite a clear physical meaning. The function $\chi (s)$ originates
from initial and final states interactions and  describes the effect
of screening the three-body forces by two-body ones \cite{25}.

If we take the usual parameterization for one-particle inclusive
cross-section in the region of diffraction dissociation
\be
\frac{s}{\pi}\frac{d\sigma}{dtdM_X^2} = A(s.M_X^2)\exp[b(s,M_X^2)t],
\label{32}
\ee 
then we obtain for the quantities $A$ and $b$
\be
A(s,M_X^2) = \frac{(2\pi)^3}{I(s)}\chi(\bar s)f_0(\bar
s),\quad 
b(s,M_X^2) = \frac{R_0^2(\bar s)}{2} \label{33}.
\ee

Eq. (\ref{33}) shows that the effective radius of three-body forces
is related to the slope of diffraction cone for inclusive diffraction
dissociation processes in the same way as the effective radius of
two-body forces is related to the slope of diffraction cone in
elastic scattering processes. Moreover, it follows from the
expressions
\[
R_0(\bar s) = \frac{r_0}{M_0} \ln \bar s/s'_0,\quad \bar s =
2(s+M_N^2) - M_X^2
\]
that the slope of diffraction cone for inclusive
diffraction dissociation processes at fixed energy decreases  with
the growth  of missing mass. This property agrees well qualitatively 
with the experimentally observable picture.

Hence physically tangible notion of the effective radius of
three-body forces introduced previously provides a clear physical
interpretation that helps one to create a visual picture and
representation for inclusive diffraction dissociation
processes at the same level as one can understand and represent
elastic scattering processes at high energies. Besides, 
relation (\ref{28}) together with linear equation (\ref{7}) for the
three-body forces scattering amplitude may be the basis of powerful
dynamic apparatus for constructing the dynamical models for the 
theoretical description of the inclusive reactions.

In the case of unitarity saturation of the three-body forces, we have
from generalized asymptotic theorems
\[
f_0(s) \sim  s^{3/2}\left(\frac{\ln s/s'_0}{M_0}\right)^5,\quad
\chi(s) \sim  \frac{1}{\sqrt{s}{\ln}^3 s},\quad s\rightarrow 
\infty.
\]
This means that

\be
\fbox{$\displaystyle A(s,M_X^2) \sim \ln^2\frac{\bar
s}{s'_0},\quad
s\rightarrow 
\infty$}\,.\label{34}
\ee

On the other hand, comparing formulae (\ref{25}) and (\ref{33}) we
see that one and the same combination $\chi f_0$ enters in the
equations. Therefore, we can extract this combination and express it
through experimentally measurable quantities. We have in this way
\be
A(s,M_X^2) = \frac{\bar s M_N
R_d^3}{(2\pi)^{3/2}I(s)}\delta\sigma_0(\bar s). \label{35}
\ee
In that case it would be very desirable to think about the creation
of accelerating deuterons beams instead of protons ones at the now  
working accelerators and colliders.

\section{On the structure of hadronic total cross-sections}

Let's rewrite the equation for $\chi(s)$ 
\[
\chi(s) = \frac{\sigma_{hN}^{tot}(s/2)}{2\pi[B_{hN}(s/2)+R_0^2(s)]} -
1
\]
in the form

\be
\sigma_{hN}^{tot}(s) = 2\pi\left[B_{hN}(s) +
R_0^2(2s)\right]\left(1+\chi\right).\label{36}
\ee
From the Froissart and generalized asymptotic bounds we have
\[
\chi(s) = O\left(\frac{1}{\sqrt{s}\ln^3s}\right),\quad s\rightarrow
\infty.
\]
We also know that \cite{20}
\be
\sigma_{hN}^{tot}(s,s_0) \sim \ln^2(s/s_0)\Longrightarrow
B_{hN}(s,s_0)\sim \ln^2(s/s_0),\label{37}
\ee
and Eq. (\ref{36}) gives
\[
R_0^2(2s,s'_0)\sim \ln^2(2s/s'_0)\sim \ln^2(s/s_0),\quad
s\rightarrow\infty.
\]
Therefore, we come to the following asymptotic consistency condition:
\be
\fbox{$\displaystyle s'_0 = 2s_0$}\,.\label{38}
\ee
The asymptotic consistency condition tells us that we have not any
new scale. The scale defining unitarity saturation of three-body
forces is unambiguously expressed by the scale which defines
unitarity saturation of two-body forces. In that case we have
\[
R_0^2(2s,s'_0) = R_0^2(s,s_0)
\]
and
\be
\fbox{$\displaystyle \sigma_{hN}^{tot}(s) = 2\pi\left[B_{hN}(s) +
R_0^2(s)\right]\left(1+\chi(s)\right)$} \label{39}
\ee
with a common scale $s_0$. 

Reminding the relation between the effective radius of two-body
forces and the slope of diffraction cone in elastic scattering 
\be
B_{hN}(s) = \frac{1}{2}R_{hN}^2(s), \label{40}
\ee
we obtain
\be
\sigma_{hN}^{tot}(s) = \pi R_{hN}^2(s) +
2\pi R_0^2(s),\quad s\rightarrow\infty.\label{41}
\ee
Equations (\ref{39}) and (\ref{41}) define a new nontrivial structure
of hadronic total cross-section. It should be emphasized that the
coefficients staying in the R.H.S. of Eq. (\ref{41}) in front of
effective radii of two- and three-body forces are strongly fixed.

It is useful to compare the new structure of total hadronic
cross-section with the known structure. We have from unitarity

\be
\sigma_{hN}^{tot}(s) = \sigma_{hN}^{el}(s) +
\sigma_{hN}^{inel}(s).\label{42}
\ee
If we put
\be
\sigma_{hN}^{el}(s) = \pi R_{hN}^{el\,^2}(s),\quad
\sigma_{hN}^{inel}(s) = 2\pi R_{hN}^{inel\,^2}(s),\label{43}
\ee
then we come to the similar formula

\be
\sigma_{hN}^{tot}(s) = \pi R_{hN}^{el\,^2}(s) + 2\pi
R_{hN}^{inel\,^2}(s).\label{44}
\ee
But it should be borne in mind
\be
R_{hN}^2(s) \not= R_{hN}^{el\,^2}(s),\quad R_0^2(s) \not=
R_{hN}^{inel\,^2}(s).\label{45}
\ee
In fact, we have
\be
\sigma_{hN}^{el}(s) = \frac{\sigma_{hN}^{tot\,^2}(s)}{16\pi
B_{hN}(s)}
= \frac{\sigma_{hN}^{tot\,^2}(s)}{8\pi R_{hN}^2(s)},\label{46}
\ee
\be
\sigma_{hN}^{inel}(s) = \sigma_{hN}^{tot}(s)\left[1 -
\frac{\sigma_{hN}^{tot}(s)}{8\pi R_{hN}^2(s)}\right].\label{47}
\ee
Of course, Eqs. (\ref{43}) are the definitions of $R_{hN}^{el}$ and
$R_{hN}^{inel}$. The definition of $R_{hN}^{el}$ corresponds to our
classical imagination, the definition of $R_{hN}^{inel}$
corresponds to our knowledge of quantum mechanical problem for
scattering from the black disk. Let us suppose that
\be
\sigma_{hN}^{tot}(s_m) \cong  \pi R_{hN}^2(s_m),\ \
s_m\in{\cal M},\quad \left(R_0^2(s_m)\ll
R_{hN}^2(s_m)\right),\label{48}
\ee
then we obtain
\be
\sigma_{hN}^{el}(s_m) = \frac{1}{8}\pi R_{hN}^2(s_m),\quad
\sigma_{hN}^{inel}(s_m) = \frac{7}{8}\pi R_{hN}^2(s_m).\label{49}
\ee
This simple example shows that the new structure of total hadronic
cross-sections is quite different from that given by Eq.
(\ref{42}). The reason is that the structure (\ref{39}) is of the
dynamical origin. We have mentioned above that the coefficients,
staying in the R.H.S. of Eq. (\ref{41}) in front of effective radii
of two- and three-body forces, are strongly fixed. In fact, we found
here the answer to the old question: Why the constant ($\pi/m_{\pi}^2
\approx 60\,mb $) staying in the Froissart bound is too large in the
light of the existing experimental data. The constant in the R.H.S.
of Eq. (\ref{41}), staying in front of effective radius of
hadron-hadron interaction, is 4 times smaller than the constant in
the Froissart bound. But this is too small to correspond to the
experimental data. The second term in the R.H.S. of Eq. (\ref{41})
fills an emerged gap.

It is a remarkable fact that the quantity $R_0^2$, which has the
clear physical interpretation, at the same time, is related to the
experimentally measurable quantity which the total cross-section is.
This important circumstance gives rise to the new nontrivial
consequences which are discussed in the next section.

We made an attempt to check up the structure (\ref{39}) on its
correspondence to the existing experimental data and I'd like to
present the preliminary results here.

At the first step, we made a weighted fit to the experimental data on
the proton-antiproton total cross-sections in the range 
$\sqrt{s}>10\, GeV$. The data were fitted with the function of the
form predicted by Froissart bound in the spirit of our
approach\footnote{Recently, from a careful analysis of the
experimental data and a comparative study of the known characteristic
parameterizations, Bueno and Velasco have shown (Phys. Lett. B{\bf
380}, 184 (1996)) that statistically a ``Froissart-like" type
parameterization for proton-proton and proton-antiproton total
cross-sections is strongly favoured.}
\be
\sigma^{tot}_{asmpt} = a_0 + a_2 \ln^2(\sqrt{s}/\sqrt{s_0})
\label{50}
\ee
where $a_0, a_2, \sqrt{s_0}$ are free parameters. We accounted for
experimental errors $\delta x_i$ (statistical and systematic errors
added in quadrature) by fitting to the experimental points with the
weight $w_i=1/(\delta x_i)^2$. Our fit yielded
\be
a_0 = (42.0479\pm 0.1086)mb,\quad a_2 = (1.7548\pm
0.0828)mb,\label{51}
\ee
\be
\sqrt{s_0} = (20.74\pm 1.21)GeV.\label{52}
\ee
The fit result is shown in Fig. 3.

After that we made a weighted fit to the experimental data on the
slope of diffraction cone in elastic $p\bar p$ scattering. The
experimental points and the references, where they have been
extracted from, are listed in  \cite{26}. The fitted function of
the form
\be
B = b_0 + b_2 \ln^2(\sqrt{s}/20.74), \label{53}
\ee
which is also suggested by the asymptotic theorems of local quantum
field theory, has been used. The value $\sqrt{s_0}$ has been fixed by
(\ref{52}) from the fit to the $p\bar p$ total cross-sections data.
Our fit yielded
\be
b_0 = (11.92\pm 0.15)GeV^{-2} ,\quad b_2 = (0.3036\pm
0.0185)GeV^{-2}.\label{54} 
\ee
The fitting curve is shown in Fig. 4.

At the final stage we build a global (weighted) fit to all the data
on proton-antiproton total cross-sections in a whole range of
energies available up today. The global fit was made with the
function of the form
\be
\sigma^{tot}_{p\bar p}(s) = \sigma^{tot}_{asmpt}(s) \left[1 +
\frac{c}{\sqrt{s-4m^2_N}R^3_0(s)} \left(1+\frac{d_1}{\sqrt{s}} +
\frac{d_2}{s} + \frac{d_3}{s^{3/2}}\right)\right] \label{55}
\ee
where $m_N$ is proton (nucleon) mass,
\be
R^2_0(s) = \left[0.40874044 \sigma^{tot}_{asmpt}(s)(mb) -
B(s)\right](GeV^{-2}),\label{56}
\ee
\be
\sigma^{tot}_{asmpt}(s) = 42.0479 + 1.7548
\ln^2(\sqrt{s}/20.74),\label{57}
\ee
\be
B(s) = 11.92 + 0.3036 \ln^2(\sqrt{s}/20.74),\label{58}
\ee
$c, d_1, d_2, d_3$ are free parameters. Function (\ref{55})
corresponds to the structure given by Eq. (\ref{39}). 

In fact, we have for the function $\chi (s)$ in the R.H.S.
of Eq. (\ref{39}) theoretical expression in the form
\be
\chi (s) = \frac{C}{\kappa (s)R_0^3(s)} \label{59}
\ee
where
\be
\kappa^4 (s) = \frac{1}{2\pi}\int_a^b dx
\sqrt{(x^2-a^2)(b^2-x^2)[(a+b)^2-x^2]},\label{60}
\ee
\[
a = 2m_N,\quad b = \sqrt{2s+m_N^2}-m_N.
\]
It can be proved that $\kappa (s)$ has the following
asymptotics\footnote{Integral in R.H.S. of Eq.(\ref{60}) can be
expressed in terms of the Appell function.}
\[
\kappa (s)\sim \sqrt{s},\quad s\rightarrow \infty,
\]
\[
\kappa (s)\sim \sqrt{s-4m_N^2},\quad s\rightarrow 4m_N^2.
\]
We used at the moment the simplest function staying in the R.H.S. of
Eq. (\ref{55}) which described these two asymptotics.

Our fit yielded
\[
d_1 = (-12.12\pm 1.023)GeV,\quad d_2 = (89.98\pm 15.67)GeV^2,
\]
\be
d_3 = (-110.51\pm 21.60)GeV^3,\quad c = (6.655\pm
1.834)GeV^{-2}.\label{61}
\ee
The fitting curve is shown in Figs. 5, 6.

The experimental data on proton-proton total cross-sections display a
more complex structure at low energies than the proton-antiproton
ones. To describe this complex structure we, of course, have to
modify formula (\ref{55}) without destroying the general structure
given by Eq. (\ref{39}). The modified formula looks like
\[
\sigma_{pp}^{tot}(s) = \sigma^{tot}_{asmpt}(s) \times
\]
\be
\left[1 + \left(\frac{c_1}{\sqrt{s-4m^2_N}R^3_0(s)} -
\frac{c_2}{\sqrt{s-s_{thr}}R^3_0(s)}\right)\left(1 + d(s)\right) +
Resn(s)\right],\label{62}
\ee
where $\sigma^{tot}_{asmpt}(s)$ is the same as in proton-antiproton
case (Eq. (\ref{57})) and
\be
d(s) = \sum_{k=1}^{8}\frac{d_k}{s^{k/2}},\quad Resn(s) =
\sum_{i=1}^{8}\frac{C_R^i s_R^i
{\Gamma_R^i}^2}{\sqrt{s(s-4m_N^2)}[(s-s_R^i)^2+s_R^i{\Gamma_R^i}^2]}.
\label{63}
\ee
Compared to Eq. (\ref{55}) we introduced here an additional term
$Resn(s)$ describing diproton resonances which have been extracted
from  \cite{27,28}. The positions of resonances and their
widths are listed in Table I. The $c_1, c_2, s_{thr}, d_i, C_R^i
(i=1,...,8)$ were considered as free fit parameters. The fitted
parameters obtained by  fit are listed below (see $C_R^i$ in Table
I.)
\[
c_1=(192.85\pm 1.68) GeV^-2,\quad c_2=(186.02\pm 1.67) GeV^-2,
\]
\[
s_{thr}=(3.5283\pm 0.0052) GeV^2,
\]
\[
d_1=(-2.197\pm 1.134)10^2 GeV,\quad d_2=(4.697\pm 2.537)10^3 GeV^2,
\]
\[
d_3=(-4.825\pm 2.674)10^4 GeV^3,\quad d_4=(28.23\pm 15.99)10^4 GeV^4,
\]
\[
d_5=(-98.81\pm 57.06)10^4 GeV^5,\quad d_6=(204.5\pm 120.2)10^4 GeV^6,
\]
\be
d_7=(-230.2\pm 137.3)10^4 GeV^7,\quad d_8=(108.26\pm 65.44)10^4
GeV^8.\label{64}
\ee
The fitting curve is shown in Figs. 7-10. It should be pointed out
that our fit revealed that the resonance with the mass $m_R=2106\,
MeV$ should be odd parity. Our fit indicates that this resonance is
strongly confirmed by the set of experimental data on proton-proton
total cross-sections. That is why a further study of diproton
resonances is very desirable. 

The figures 4-11 display a very good correspondence of theoretical
formula (\ref{39}) to the existing experimental data on proton-proton
and proton-antiproton total cross-sections. 

I'd like to emphasize the following attractive features of formula
(\ref{39}). This formula represents hadronic total cross-section in a
factorized form. One factor describes high-energy asymptotics of
total cross-section and it has the universal energy dependence
predicted by the Froissart theorem. Other factor is responsible for
the behaviour of total cross-section at low energies and this factor
has also a universal asymptotics at the  elastic threshold. It is a
remarkable fact that the low-energy asymptotics of total
cross-section at the elastic threshold is dictated by the high-energy
asymptotics of three-body (three-nucleon in that case) forces. 
This means that we undoubtedly faced very deep physical
phenomena here. The appearance of new threshold $s_{thr}=3.5283\,
GeV^2$ in proton-proton channel, which is near to the elastic
threshold, is a nontrivial fact too. It's clear that the difference
of two identical terms with different thresholds in the R.H.S. of Eq.
(\ref{62}) is a tail of crossing symmetry which was not actually
taken into account in our consideration. What physical entity does
this new threshold correspond  to? This interesting question is still
open. 

Anyway we have established that simple theoretical formula (\ref{39})
described the global structure of $pp$ and $p\bar p$ total
cross-sections in the whole range of energies available up today. Of
course, our results concerning a global description of hadronic total
cross-sections are to be considered as preliminary ones. We know the
ways how they can be refined later on.

\section{On the slope of diffraction cone in single diffraction
dissociation}

We have shown above that the slope of diffraction cone in the single
diffraction dissociation is related to the effective interaction
radius for three-body forces
\be
b_{SD}(s,M_X^2) = \frac{1}{2}R_0^2(\bar s,s'_0),\label{65}
\ee
\[
\bar s = 2(s + M_N^2) - M_X^2, \quad s'_0 = 2s_0.
\]

Let us define the slope of diffraction cone in the single diffraction
dissociation at a fixed point over the missing mass
\be
B^{sd}(s) = b_{SD}(s,M_X^2)\vert_{M_X^2 = 2M_N^2}.\label{66}
\ee
Now taking into account Eqs. (\ref{40},\ref{41}) where the effective
interaction radius for three-body forces can be extracted from
\be
R_0^2(2s,2s_0) = R_0^2(s,s_0) = \frac{1}{2\pi}\sigma^{tot}(s) -
B^{el}(s),\label{67}
\ee
and the equation
\be
\sigma^{el}(s) = \frac{\sigma^{tot\,^2}(s)}{16\pi B^{el}(s)},\quad
(\rho = 0)\label{68}
\ee
we come to the fundamental relation between the slopes in the single
diffraction dissociation and elastic scattering
\be
\fbox{$\displaystyle B^{sd}(s) = B^{el}(s)\left(4X -
\frac{1}{2}\right)$}\,,\label{69}
\ee
where
\be
X \equiv \frac{\sigma^{el}(s)}{\sigma^{tot}(s)}.
\ee
The quantity $X$ has a clear physical meaning, it has been introduced
in the papers of C.N. Yang and his collaborators \cite{29,30}.

We found $X = 0.25$ at $\sqrt{s} = 1800\, GeV$ (see the CDF paper
mentioned in Introduction). Hence, in that case we have  $B^{sd} =
B^{el}/2$ which is confirmed not so badly in the experimental
measurements.

In the limit of the black disk $(X = 1/2)$ we obtain
\be
\fbox{$\displaystyle B^{sd} = \frac{3}{2}B^{el}$}\,,\label{70}
\ee
and
\be
\fbox{$\displaystyle B^{sd} = B^{el},\quad at\quad X = \frac{3}{8} =
0.375$}\,.
\ee

So, we observe that there is quite a nontrivial dynamics in the
slopes of diffraction cone in the single diffraction dissociation and
elastic scattering processes. In particular, we can study an
intriguing question on the black disk limit not only in the
measurements of total hadronic cross-sections compared with elastic
ones but in the measurements of the slopes in single diffraction
dissociation processes together with  elastic scattering ones. 

There is a more general formula which can be derived with account of
the real parts for the amplitudes. This formula  looks like
\be
\fbox{$\displaystyle B^{sd}(s) = B^{el}(s)\left(4X\frac{1 -
\rho_{el}(s)\rho_0(s)}{1 + \rho_{el}^2(s)} -
\frac{1}{2}\right)$}\,.\label{71}
\ee
If $\rho_{el} = 0$ or $\rho_0 = -\rho_{el}$, then we come to 
Eq. (\ref{69}). In the case when $\rho_{el} \not= 0$, we can rewrite
Eq. (\ref{71}) in the form
\be
\rho_0 = \frac{1}{\rho_{el}}\left[1 - \frac{1 +
\rho_{el}^2}{8X}\left(1 +
\frac{2B^{sd}}{B^{el}}\right)\right].\label{72}
\ee
Eq. (\ref{72}) can be used for the calculation of the new quantity
$\rho_0$. Anyway,  the measurements of real parts for the
amplitudes seem to play an important role in the future high energy
hadronic physics.

\section{On total single diffractive dissociation\\ cross-section}

For the total single diffractive dissociation cross-section defined
as
\be
\sigma_{sd}(s) =
2\pi\int_{M_{min}^2}^{0.1s}\frac{dM_X^2}{s}\int_{t_{-}(M_X^2)}^{t_{+}
(M_X^2)} dt
A(s,M_X^2)\exp[b(s,M_X^2)t] \label{73}
\ee
we obtained the following asymptotic formula
\be
\sigma_{sd}(s) = \frac{A_0 + A_2\ln^2(\sqrt{s}/\sqrt{s_0})}{c_0 +
c_2\ln^2(\sqrt{s}/\sqrt{s_0})},\label{74}
\ee
where $c_0, c_2$ are related to effective interaction radius for
three-body forces
\[
R_0^2(s) = c_0 + c_2\ln^2(\sqrt{s}/\sqrt{s_0}),
\]
and $A_0, A_2$ to be found from the fit to the experimental data on
$\sigma_{sd}$. The experimental values for $p\bar p$ single
diffraction dissociation cross-sections, which were used, are listed
in Table II. Our fit yielded \cite{26,31}
\[
A_0 = 23.395\pm 2.664\, mbGeV^{-2},\quad A_2 = 4.91\pm 0.26\,
mbGeV^{-2}.
\]
The fit result is shown in Fig. 11. As you can see, the fitting curve
goes excellently over the experimental points of the CDF group at
Fermilab.

Thus, we have shown that from the generalized asymptotic theorems a
l\`a Froissart there follows  a simple formula which allows
one to match the experimental data on $p\bar p$ single diffraction
dissociation cross-sections at high energies including lower energies
as well. At present only the suggested approach allows one to
quantitatively describe the observed behaviour of $p\bar p$ single
diffraction dissociation cross-sections.

Some time ago many high energy physicists thought that the increase
of total cross-sections was due to the same increase of single
diffraction dissociation cross-sections. Now we know that this
thought is wrong and, moreover, we understand why this is the case.

As it has been shown above the phenomenon of exceedingly moderate
energy dependence of single diffraction dissociation cross-sections
on $s$ observed by CDF at Fermilab is a manifestation of unitarity
saturation of three-nucleon forces at Fermilab Tevatron energies.
This phenomenon is confirmed in the dynamics consistent with
unitarity becoming apparent in the effect of screening of three-body
forces by two-body ones. It is to be compared with the discovery
of the increase of $pp$ total cross-sections at CERN ISR and of the
growth of $K^{+}p$ total cross-sections revealed at Serpukhov
accelerator. In this context, the CDF data are the ones of the most
significant experimental results obtained in the last years.

In fact, we have found the bound (like Froissart bound!)
\be
\fbox{$\displaystyle\sigma_{sd}(s) < Const,\quad
s \rightarrow \infty$}\,.\label{75}
\ee

I'd like especially to point out that analyticity and unitarity
together with the dynamic apparatus of single-time formalism in QFT
provide the clear answers to the asymptotic behaviour both the
elastic scattering and single diffraction dissociation at high
energies, which correspond to the experimentally observable picture.

It is very nice that the understanding of ``soft" physics based
on general principles of QFT, such as analyticity and unitarity, is
so fine confirmed by the experimentally observable picture compared
to the models where the general principles have been broken down.

I hope that it will be possible to test the obtained results at
higher energies, such as those of the LHC collider and even 
higher ones.

\section{On the forms of strong interaction\\ dynamics}

Conditionally there are two forms of strong interaction dynamics:
t-channel form and s-channel one.
\vskip 0.1 true in
\centerline{\bf t-channel form}
\vskip 0.1 true in
\noindent The fundamental  quantity here is some set of Regge
trajectories:
\be
t-channel form \qquad \Longleftrightarrow \qquad
\alpha_R(t).\label{76}
\ee
Here subscript $R$ enumerates different Regge trajectories which are
the poles in the t-channel partial wave amplitudes for the given
process. There are a lot of people who work in the field of t-channel
dynamics of strong interactions.

Some part of scientific community works in the field of s-channel
form of strong interaction dynamics.
\vskip 0.1 true in
\centerline{\bf s-channel form}
\vskip 0.1 true in
\noindent The fundamental  quantity here is an effective interaction
radius of fundamental forces:
\be
s-channel form \qquad \Longleftrightarrow \qquad
R_{\alpha}(s).\label{77}
\ee
Here subscript $\alpha$ enumerates different types of hadrons and
fundamental forces acting between them. The s-channel form of
dynamics allows one to create a physically transparent and visual
geometric picture of strong interactions for hadrons. I'd like to
emphasize the attractive features of this form of strong interaction
dynamics.
\vskip 0.1 true in
\begin{itemize}
\item Universality (existence of pion with $m_{\pi}\not=0$):
\[
\fbox{$\displaystyle R_{\alpha}(s) \sim \frac{r_{\alpha}}{m_{\pi}}\ln
\frac{s}{s_0},\quad s \rightarrow \infty$}\,.
\]
\item Compatibility with the general principles of relativistic
quantum theory.
\item Fine mathematical structures are given by the global
analyticity together with single-time formalism in QFT.
\end{itemize}
That is why, in our opinion, the s-channel form of strong interaction
dynamics is more preferable than the t-channel one. 

\section{Conclusion}

In Commemoration of the 200th Anniversary of Alexander S. Pushkin I'd
like to conclude my talk with the Ode:
\vskip 0.3 true in
\font\cyr wncyr10 at 12.0pt
{\bf
\Large
\leftline{\it To learn or not to learn?}
\leftline{\it Of course to learn to be The Forces,}
\leftline{\it To be The Three-Body Forces as well,}
\leftline{\it But not \fbox{\rm {\cyr Pomer}-{\cyr On}} alone.}
\leftline{\it That is a sound of Bell!...}}
\vskip 0.3 true in
\section*{Acknowledgements}
It is my pleasure to express thanks to the Organizing 
Committee for the given opportunity to attend the VIIIth Blois
Workshop and present the report there. I am indebted to V.V. Ezhela
for the access to the computer readable files on total proton-proton
and proton-antiproton cross-sections in the IHEP COMPAS database. The
friendly encouragement and many pieces of good advice on computer
usage from A.V. Razumov are gratefully acknowledged.

\newpage
\begin{center}
\begin{picture}(288,388)
\put(-50,-110){\includegraphics[width=14cm]{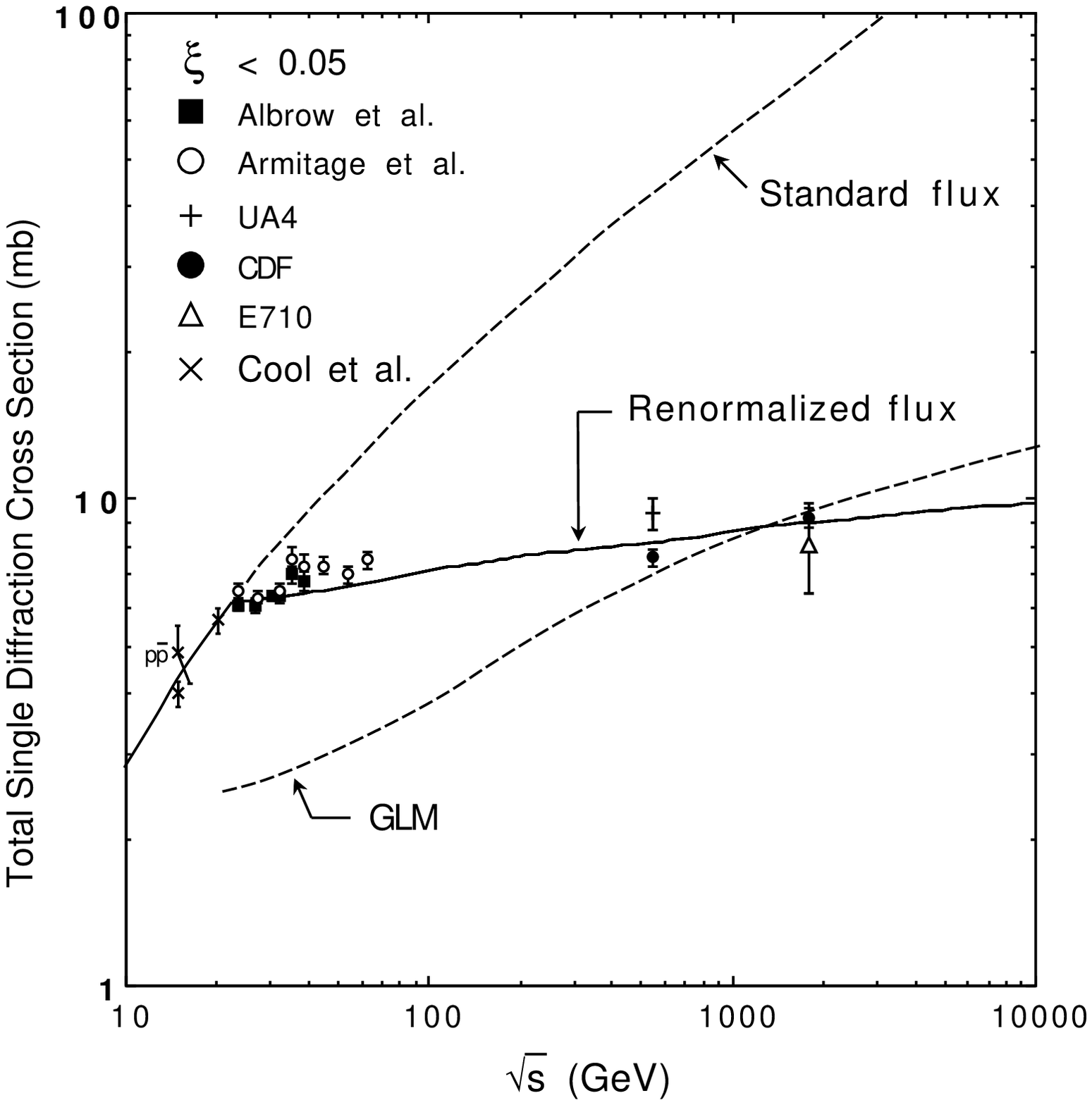}}
\end{picture}
\end{center}

Figure 1: The total single diffraction cross-sections for $p(\bar
p)+p\rightarrow p(\bar p)+X$ vs $\sqrt{s}$ compared with the
predictions of the renormalized Pomeron flux model of Goulianos
\cite{3} (solid line) and of the model Gostman, Levin and Maor
\cite{4} (dashed line, labelled GLM).
\newpage
\vspace*{-1cm}
{\Large
\[
\frac{s}{\pi}\frac{d\sigma_{hN\rightarrow NX}}{dtdM_X^2} =
\frac{(2\pi)^3}{I(s)}\chi(\bar s)Im{\cal F}_0(\bar
s;-\vec{\Delta}, \vec{\Delta}, \vec q; 
\vec{\Delta}, -\vec{\Delta}, \vec q\,)
\]
\[
\bar s = 2(s + M_N^2) - M_X^2,\quad t = - 4{\vec\Delta}^2.
\]
\centerline{($I(s)/\chi(\bar s)$ -- "renormalized flux"!)}
}
\begin{center}
\begin{picture}(275,315)
\put(-40,-50){\includegraphics[width=12cm]{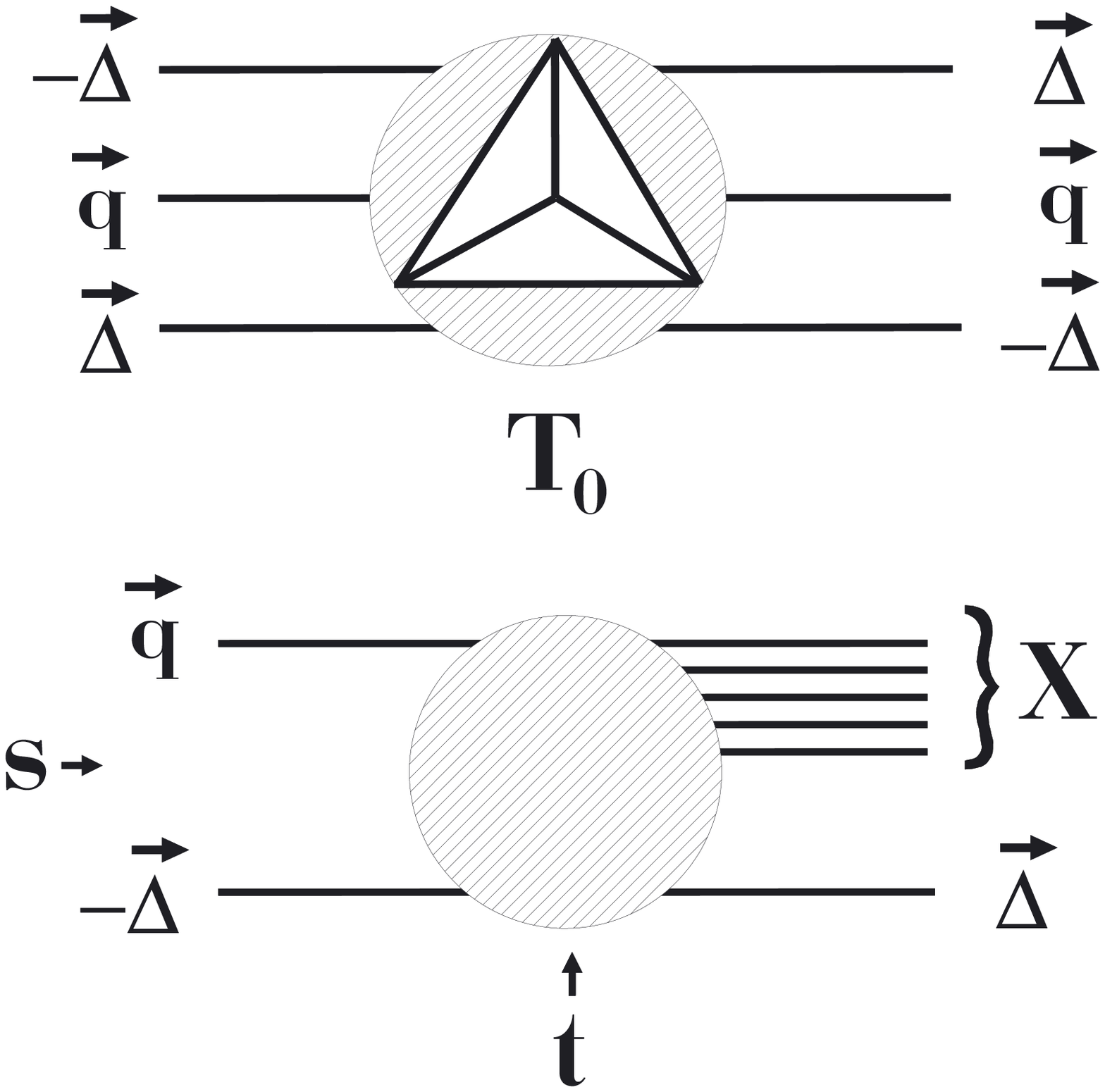}}
\end{picture}
\end{center}

\vspace{2.5cm}
Figure 2: Kinematical notations and configuration of momenta in the
relation of one-particle inclusive cross-section to the three-body
forces scattering amplitude.

\newpage
\begin{center}
\begin{picture}(288,198)
\put(15,10){\includegraphics{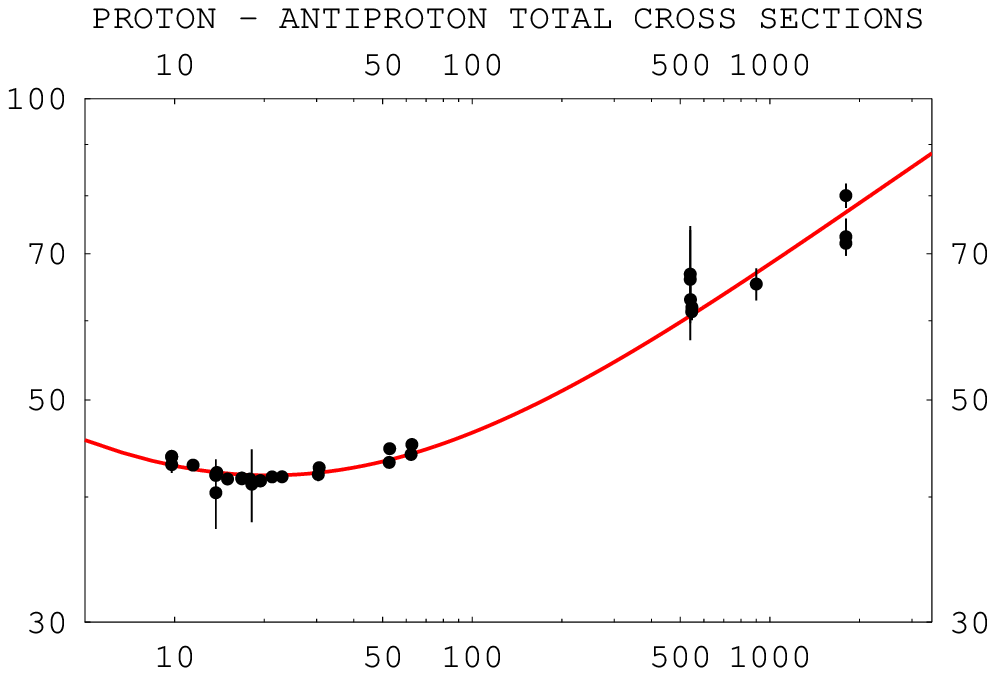}}
\put(144,0){$\sqrt{s}\, (GeV)$}
\put(0,77){\rotatebox{90}{\large$\sigma_{tot} (mb)$}}
\end{picture}
\end{center}

Figure 3: The total proton-antiproton cross-section versus
$\sqrt{s}$
compared with formula (\ref{50}). Solid line represents our fit
to the data. Statistical and systematic errors added in quadrature.

\vspace{1cm}
\begin{center}
\begin{picture}(288,188)
\put(15,10){\includegraphics{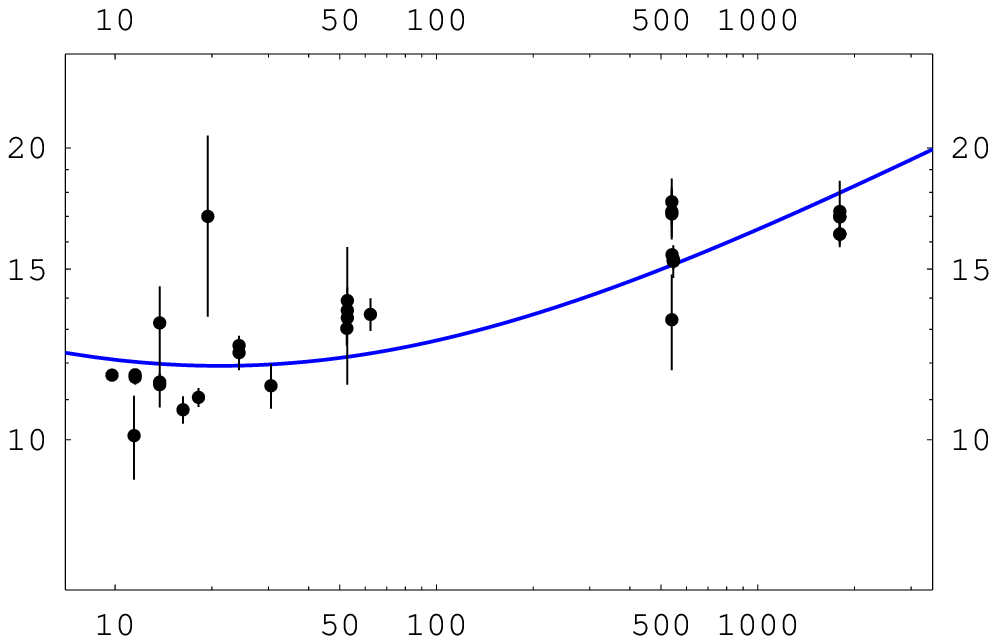}}
\put(144,0){$\sqrt{s}\, (GeV)$}
\put(0,77){\rotatebox{90}{$B\ \ (GeV^{-2})$}}
\end{picture}
\end{center}

Figure 4: Slope $B$ of diffraction cone in $p\bar p$
elastic scattering. Solid line represents our fit to the data.

\newpage
\begin{center}
\begin{picture}(288,204)
\put(15,10){\includegraphics{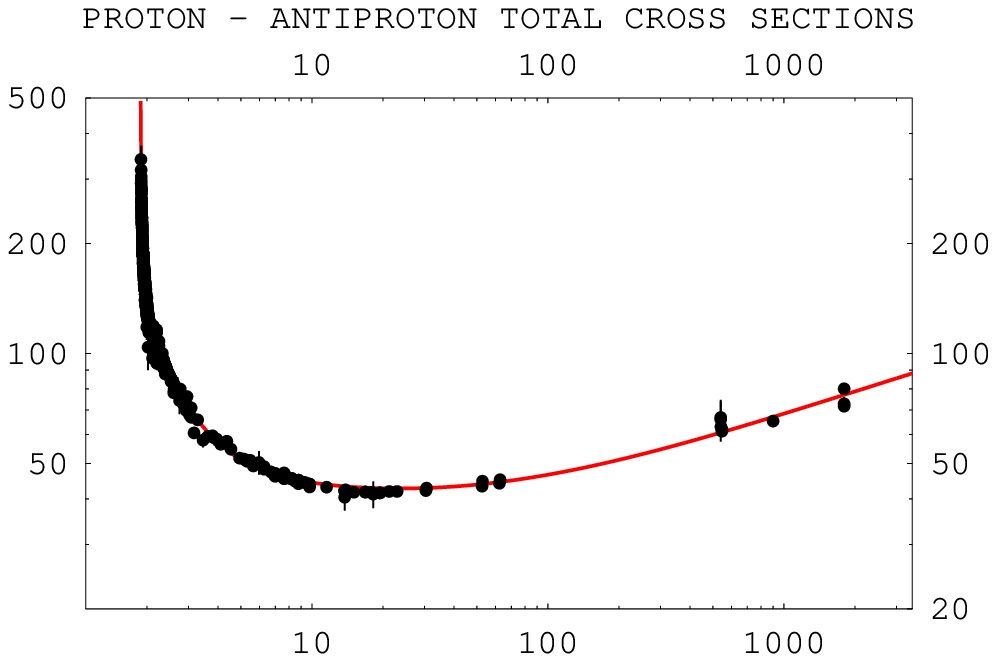}}
\put(144,0){$\sqrt{s}\, (GeV)$}
\put(0,87){\rotatebox{90}{\large$\sigma_{tot} (mb)$}}
\end{picture}
\vspace{1cm}
\end{center}

Figure 5: The total proton-antiproton cross-section versus
$\sqrt{s}$
compared with formula (\ref{55}). Solid line represents our fit
to the data. 
\vspace{1cm}
\begin{center}
\begin{picture}(288,194)
\put(15,10){\includegraphics{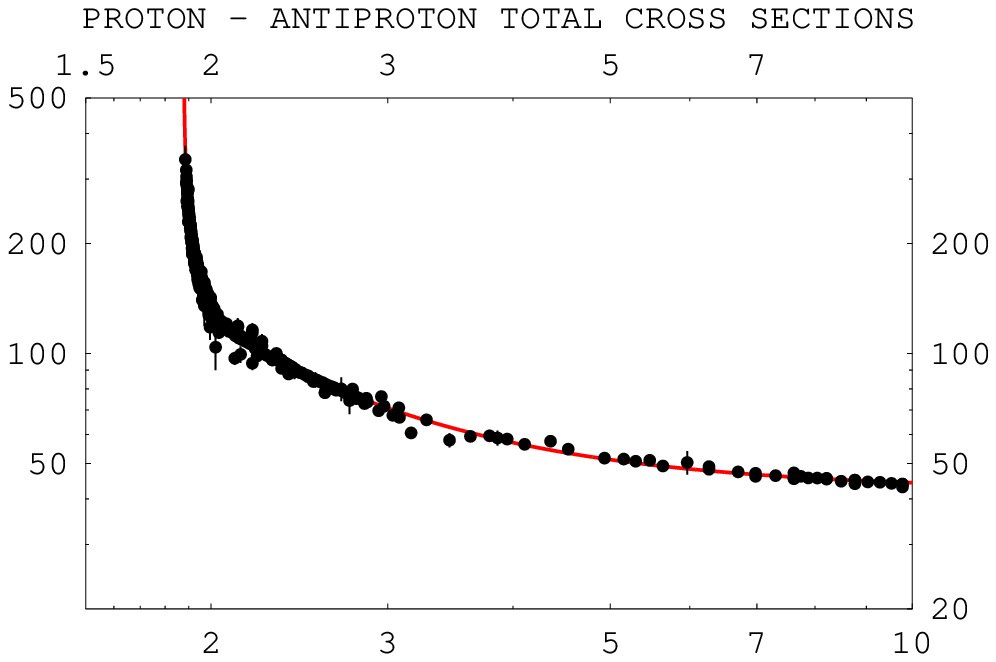}}
\put(144,0){$\sqrt{s}\, (GeV)$}
\put(0,87){\rotatebox{90}{\large$\sigma_{tot} (mb)$}}
\end{picture}
\end{center}

Figure 6: The total proton-antiproton cross-section versus
$\sqrt{s}$
compared with formula (\ref{55}) in the range $\sqrt{s}<10\, GeV$
(fragment of Fig. 5). Solid line represents our fit to the data. 

\newpage
{\large
\begin{center}
\begin{tabular}{|c|c|c|}\hline   
$ m_R(MeV) $ & $\Gamma_R(MeV) $ & $C_R(GeV^2)$  \\ \hline     
$ 1937\pm 2 $ & $ 7\pm 2 $ & $ 0.0722\pm 0.0235 $ \\ \hline
$ 1955\pm 2 $ & $ 9\pm 4 $ & $ 0.1942 \pm 0.0292 $ \\ \hline
$ 1965\pm 2 $ & $ 6\pm 2 $ & $ 0.1344 \pm 0.0117 $ \\ \hline
$ 1980\pm 2 $ & $ 9\pm 2 $ & $ 0.3640 \pm 0.0654 $ \\ \hline
$ 2008\pm 3 $ & $ 4\pm 2 $ & $ 0.3234 \pm 0.0212 $ \\ \hline
$ 2106\pm 2 $ & $11\pm 5 $ & $-0.2958 \pm 0.0342 $ \\ \hline
$ 2238\pm 3 $ & $22\pm 8 $ & $ 0.4951 \pm 0.0559 $ \\ \hline
$ 2282\pm 4 $ & $24\pm 9 $ & $ 0.0823 \pm 0.0319 $ \\ \hline
\end{tabular}
\end{center}

Table I: Diproton resonances extracted from  \cite{27,28}.

\vspace{1cm}
\begin{center}
\begin{tabular}{|c|c|c|}\hline$\sqrt{s}\ (GeV)$ & $\sigma^{p\bar
p}_{sd}(mb)$ & References \\ \hline
    20      & $ 4.9   \pm 0.55   $ & \cite{1} \\ \hline
    200     & $ 4.8  \pm  0.9    $ & \cite{32} \\ \hline
    546     & $ 5.4  \pm  1.1    $ & \cite{33} \\ \hline
    546     & $ 7.89  \pm  0.33  $ & \cite{1}  \\ \hline
    546     & $ 9.4  \pm  0.7    $ & \cite{34} \\ \hline
    900     & $ 7.8   \pm  1.2   $ & \cite{34} \\ \hline
    1800    & $ 9.46  \pm  0.44  $ & \cite{1} \\ \hline
    1800    & $ 11.7  \pm  2.3   $ & \cite{35} \\ \hline
    1800    & $ 8.1  \pm  1.7    $ & \cite{35} \\ \hline
\end{tabular}
\end{center}

Table II: Data on $p\bar p$ single diffraction dissociation
cross-sections.}

\newpage
\begin{center}
\begin{picture}(288,204)
\put(15,10){\includegraphics{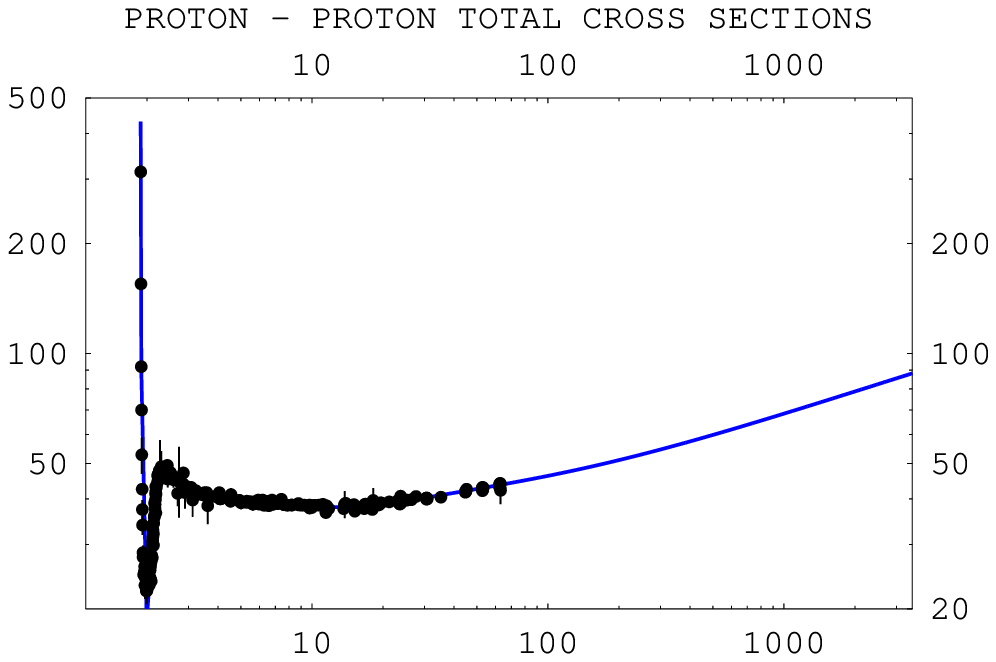}}
\put(144,0){$\sqrt{s}\, (GeV)$}
\put(0,87){\rotatebox{90}{\large$\sigma_{tot} (mb)$}}
\end{picture}
\end{center}

Figure 7: The total proton-proton cross-section versus $\sqrt{s}$
compared with formula (\ref{62}). Solid line represents our fit
to the data. Statistical and systematic errors added in quadrature.
\vspace{1cm}
\begin{center}
\begin{picture}(288,194)
\put(15,10){\includegraphics{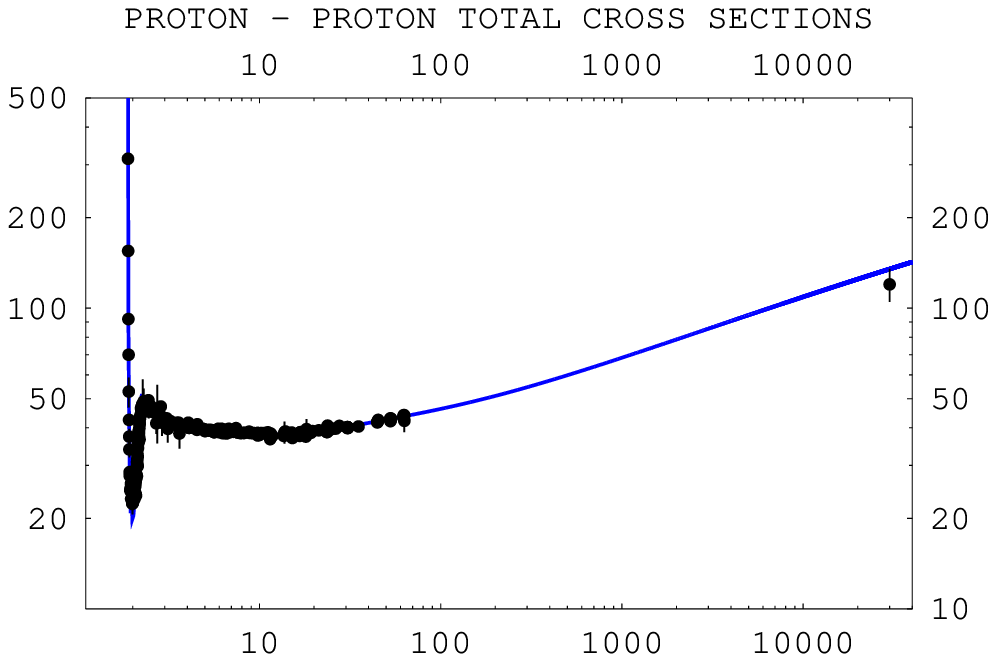}}
\put(144,0){$\sqrt{s}\, (GeV)$}
\put(0,87){\rotatebox{90}{\large$\sigma_{tot} (mb)$}}
\end{picture}
\end{center}

Figure 8: The total proton-proton cross-section (vs $\sqrt{s}$)
including a point from cosmic rays experiment \cite{36} compared with 
formula (\ref{62}). Solid line represents our fit to the data.

\newpage
\begin{center}
\begin{picture}(288,194)
\put(15,10){\includegraphics{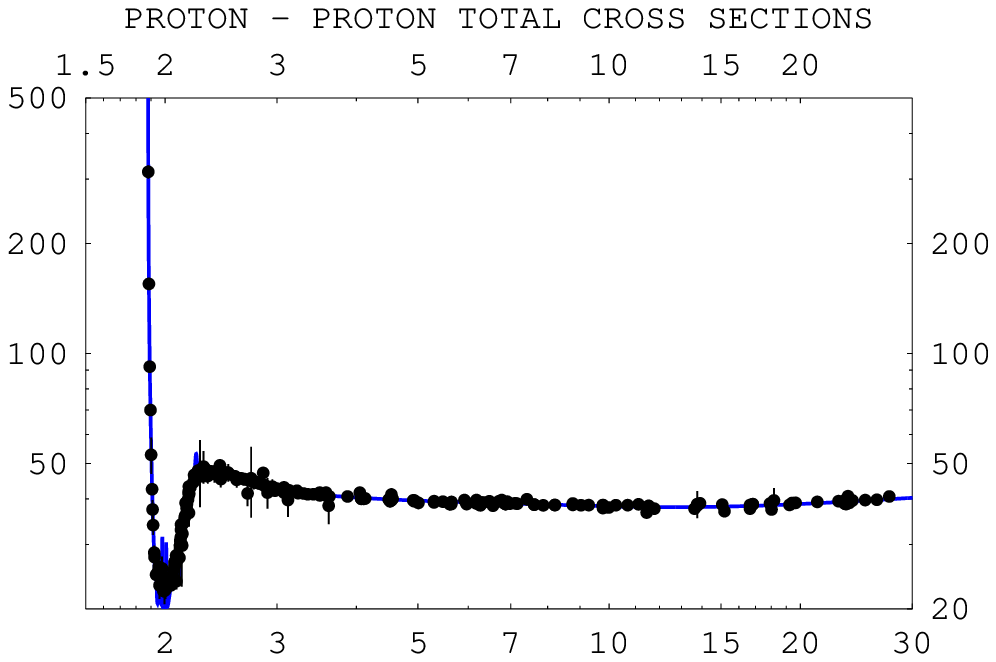}}
\put(144,0){$\sqrt{s}\, (GeV)$}
\put(0,87){\rotatebox{90}{\large$\sigma_{tot} (mb)$}}
\end{picture}
\end{center}

Figure 9: The total proton-proton cross-section in the range
$\sqrt{s}<30\, GeV$ compared with formula (\ref{62}). Solid line
represents our fit to the data.

\vspace{1cm}
\begin{center}
\begin{picture}(288,182)
\put(15,10){\includegraphics{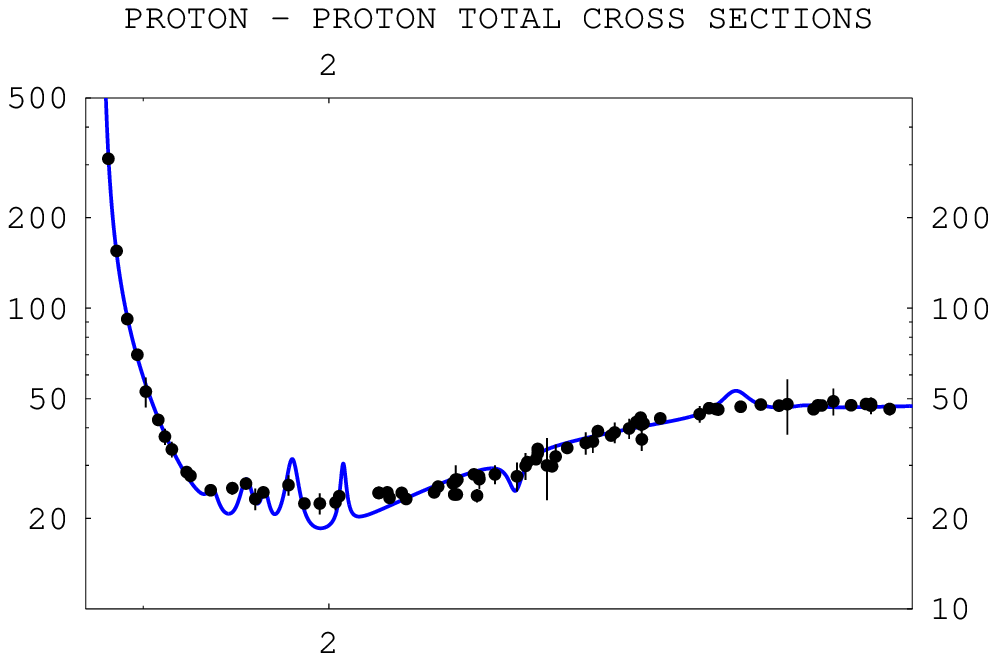}}
\put(144,0){$\sqrt{s}\, (GeV)$}
\put(0,87){\rotatebox{90}{\large$\sigma_{tot} (mb)$}}
\end{picture}
\end{center}

Figure 10: The total proton-proton cross-section at low energies
compared with formula (\ref{62}). Solid line represents our fit
to the data.

\newpage
\vspace*{2cm}
\begin{center}
\begin{picture}(388,314)
\put(18,18){\includegraphics[width=370bp]{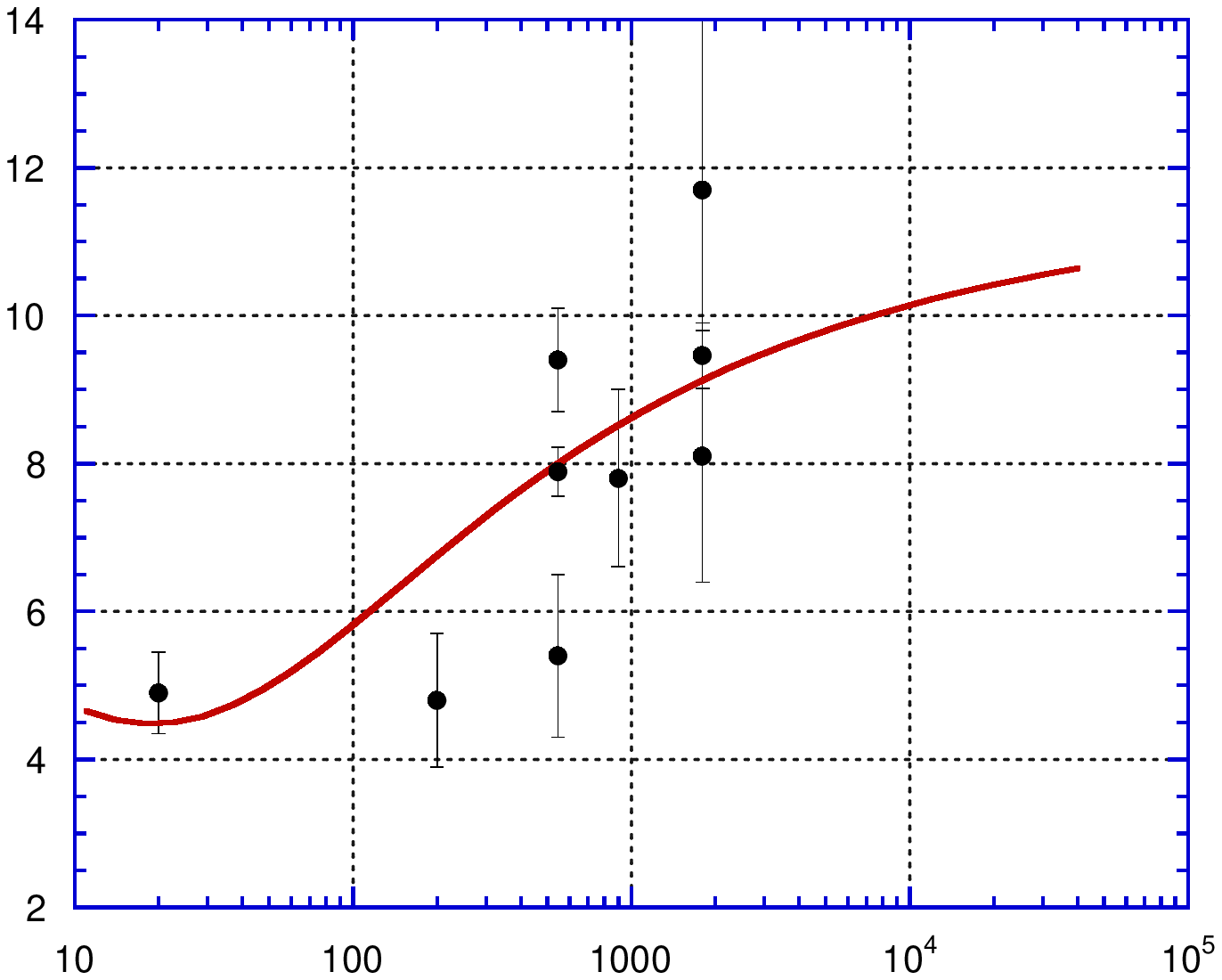}}
\put(185,0){$\sqrt{s}\, (GeV)$}
\put(0,150){\rotatebox{90}{\large$\sigma^{p\bar p}_{sd}\, (mb)$}}
\end{picture}
\end{center}

Figure 11: Total single diffraction dissociation cross-section
compared with formula (\ref{74}). Solid line represents
our fit to the data.

\end{document}